\documentclass[12pt]{article}
\usepackage{amsmath,amssymb}
\usepackage{graphicx}
\usepackage{subfig}

\newsavebox{\uuunit}
\sbox{\uuunit}
    {\setlength{\unitlength}{0.825em}
     \begin{picture}(0.6,0.7)
        \thinlines
        \put(0,0){\line(1,0){0.5}}
        \put(0.15,0){\line(0,1){0.7}}
        \put(0.35,0){\line(0,1){0.8}}
       \multiput(0.3,0.8)(-0.04,-0.02){10}{\rule{0.5pt}{0.5pt}}
     \end {picture}}

\allowdisplaybreaks

\numberwithin{equation}{section}

\begin{document}
\begin{titlepage}
\begin{center}
\vskip 3in
{\Large \textbf{A comparative study of $2d$ Ising model at different boundary
conditions using non-deterministic Hexagonal Cellular Automata}}
\vskip 8mm
{$^{\dag}$Jahangir Mohammed\footnote{
Present address: Department of Physics, Nabarangpur College, Nabarangpur-764063, Odisha, India.} and $^{*,\ddag}$Swapna Mahapatra}

{\em Department of Physics, Utkal University,\\
Bhubaneswar, Odisha 751004,
India\\
$^{\dag}$jahangirmd.physics@gmail.com\\
$^{*}$swapna.mahapatra@gmail.com\\
$^{\ddag}$swapna@iopb.res.in}
%
\end{center}

\vskip .2in
\begin{center} {\bf Abstract } \end{center}
\begin{quotation}\noindent
The spin system of the $2d$ Ising model having a hexagonal-lattice is simulated using
non-deterministic Cellular Automata. The method to implement this program is outlined
and our results show a good approximation to the exact analytic solution. The phase
transition in $2d$ Ising model is studied  with a $40\times40$ hexagonal-lattice with
five different boundary conditions (bcs) i.e., adiabatic, periodic, reflexive, fixed $+1$
and fixed $-1$ with random orientation of spins as initial conditions in the absence of an
external applied magnetic field. The critical temperature below which the spontaneous
magnetization appears as well as other physical quantities such as the magnetisation,
energy, specific heat, susceptibility and entropy with each of the bcs are calculated.
The phase transition occurs around $T^H_c$ = 1.5 which approximates well with the result
obtained from exact analytic solution by Wannier and Houtappel. We compare the behaviour
of magnetisation per cell for five different types of bcs by calculating the number of
points close to the line of zero magnetisation for $T>T^H_c$. We find that the periodic,
adiabatic and reflexive bcs give closer approximation to the  value of $T^H_c$ than fixed $+1$
and fixed $-1$ bcs with all three initial conditions for lattice size less than $50\times50$.
However, for lattice size between $50\times50$ and $200\times200$, fixed $+1$ bc and fixed $-1$
bc give closer approximation to the $T^H_c$ with initial conditions in which all spins are in
down configuration and all spins are in up configuration respectively.
\end{quotation}
\vfill
\end{titlepage}
\eject
\section{Introduction}
\label{sec:introduction}
\setcounter{equation}{0}
\par The simple Hamiltonian for the Ising model and its variants are studied extensively
for their ability to describe critical phenomena. This can be applied to understand magnetism, 
models for high-temperature superconductivity and phase diagrams, disordered and non-equilibrium 
systems. The model also provides means for testing algorithms. All the interesting phenomena of 
statistical mechanics and phase transitions are found to have counterparts in this model. It is 
an example of a lattice model for modeling simple interacting many-particle systems in statistical 
mechanics where a set of spins $s_i =\pm1$ is assigned to each lattice site.

\par The spins interact with both their nearest neighbors and an external magnetic field via 
a Hamiltonian of the form
\begin{equation}   
  H(s) = - \sum\limits_{<ij>} J_{ij} s_i s_j - \mu \sum\limits_{k=1}^N h_k s_k \label{eq:H}
\end{equation}
where $s$ is an arbitrary spin configuration, the notation $<ij>$ indicates a sum over 
nearest neighbor lattice points, $J_{ij}$($>0$) is a parameter describing inter-particle 
interactions, $\mu$ is the magnetic moment and $h_k$ is an external magnetic field at 
$k^{th}$ spin.

\par The  model is best known for describing the emergence of ferromagnetism in crystals of atoms 
that interact via spin-spin coupling. The Ising model is useful even in cases for which it
cannot be solved exactly. On both square-lattice and hexagonal-lattice geometries where the 
Ising model can be exactly solved and  also for those where  no exact solution has been found,
numerical simulations can provide insight into critical behavior of these models. The most 
common method for simulating Ising systems is the Metropolis-Hastings algorithm, originally 
developed for use in molecular dynamics simulations.

\par The $1d$ model introduced by Lenz was exactly solved by Ising, which showed that the 
one-dimensional case does not exhibit a phase transition \cite{1}. This led to the general 
belief that the Ising  model in higher dimensions would also be ineffective at describing 
systems with critical points. However, Onsager in 1944 exactly solved the Ising model for 
a two-dimensional square lattice configuration with periodic boundary condition  by analytical 
method \cite{2}. He demonstrated that, for an infinite square lattice, the ferromagnetic phase 
transition occurs at a critical temperature, $T^S_c$, of
\begin{equation} 
 { T^S_c} = \frac{2}{\ln{(1+\sqrt{2})}} \approx 2.269 \label{eq:T1}
\end{equation}
where $J=1$, $k_B = 1$ and $S$ stands for square-lattice. The square and hexagonal lattices, 
both, have exactly computable critical temperatures. The hexagonal (honeycomb) lattice is the next
simplest two dimensional lattice. The pure Ising model on the honeycomb lattice has been studied 
by Wannier \cite{3} and Houtappel \cite{4} and have been exactly solved. The most straightforward 
way to find these critical temperatures is via duality between high-temperature and low-temperature 
behaviour. Duality is the hidden symmetry found by Kramers and  Wannier that relates the partition 
function and the free energy of the Ising model at low and high temperatures for the two dimensional 
square-lattice \cite{5, 6}. The system is mathematically modeled in two ways each of which is 
independent of the other and both of which are valid to describe the physical properties of the system.
If one assumes that the free energy is singular at the critical temperature, and  that this singularity 
is unique, then this leads to the determination of the critical temperature. For the two dimensional
hexagon system the critical temperature ($T^H_c$) was found to be
\begin{equation}    
  {T^H_c} = \frac{2}{\ln{(2+\sqrt{3})}} \approx 1.519  \label{eq:T2}
\end{equation}
where $J=1$, $k_B = 1$ and $H$ stands for hexagonal-lattice. The above results are for lattice of 
infinite size with periodic boundary condition. Above this temperature the average magnetization 
is zero in the absence of external magnetic field. We performed the simulation taking lattices of
finite size and in the absence of external magnetic field so as to compare the result obtained by 
the above analytic method as well as to those for our five different bcs.

\par A Cellular Automaton (CA) is a mathematical model, modeling a set of cells which interact with 
their neighbors. In this model, each cell have values known as states, all the cells update their 
states simultaneously at discrete time steps, and the new state of a cell is determined by current 
state of its neighbors according to a local function called rule of the CA. Hexagonal CA is a 
tessellation of the plane by regular hexagons which provide for higher packing density of cells and
the unit cells of hexagonal grids are uniformly connected in the sense that the distance from a given 
cell to any adjacent cell is the same. To simulate the Ising model with Hexagonal CA, we can design a 
two states CA, for spin up state ($+1$) and spin down state ($-1$) and number of neighbors is six i.e, 
north, south, east, west, top and bottom. A definite rule is designed for which states of the cells 
are either all in up states or all in down states below $T^H_c$ and above $T^H_c$ on the average half 
in $+1$ spin states and half in $-1$ spin states. The $2d$ Ising model using non-deterministic
hexagonal CA (HCA) has not been studied for five different boundary conditions. The CA used in this 
work differs from that of Metropolis-Hastings algorithm where the algorithm specifies that the 
transitions must be made for one site at a time whereas we consider transitions of many sites 
simultaneously. The work of Eltinge \cite{7} uses Metropolis-Hastings algorithm to study numerically 
the case of periodic boundary condition.

\par Our computations will be on finite-size lattices with four different boundary conditions, 
as well as, with the  periodic boundary condition that effectively simulates accurately 
infinite-size lattice results in many respects. Apart from the Ising model, quantum spin models
such as the Kitaev model \cite{8} or Kitaev-Heisenberg model \cite{9, 10} on the honeycomb lattice 
have recently received a lot of theoretical and experimental attention \cite{11, 12, 13, 14, 15}. 
Hexagonal lattice has also been discussed in the context of high temperature and low temperature 
susceptibility series and extracting the scaling function \cite{16}. Such a growing interest in 
the hexagonal-lattice systems motivates us to revisit the Ising model on the hexagonal lattice. 
Attempts have been made for mapping Ising models in different lattice geometry using CA.
A deterministic CA (DCA) proposed by Domany and Kinzel \cite{17}, the Q2R CA \cite{18,19,20,21} 
and the Creutz CA \cite{22,23,24,25} are mostly used in analysing square-lattice Ising model. 
All these CA models are deterministic and the computation can be performed  fast. It has been 
demonstrated that the probabilistic model of the CA like Metropolis algorithm \cite{26}
is more realistic for description of the Ising model even though the random number generation 
makes it slower. Probabilistic CA model under five different boundary conditions has been 
studied in the context of a square-lattice Ising model \cite{27}. However, Ising model using 
two dimensional hexagonal CA (HCA) under different boundary conditions other than periodic 
boundary condition has not yet been studied.

\par The paper is organised as follows: in \emph{section 2}, we discuss the basic
theory to treat a $2d$ HCA and how to implement it in the Ising model with five
different bcs. The results obtained out of the simulations  are given and are analysed  in
\emph{section 3} and we have also compared Ising model with different boundary
conditions by considering their converging points. Our conclusion and future perspective
are discussed in \emph{section 4}.

\section{Implementation of Isotropic $2d$ Ising Model by Hexagonal CA}
\label{sec:cellular-automata}

\setcounter{equation}{0}

\par Two dimensional CA is described by finite states of cells ($s$), neighborhood cells 
($n$) and its distance among neighbourhood ($r$), boundary conditions and transition 
functions or rules ($f$). In our $2d$ HCA model, $s = \{s_{i,j},\, s_{i,j}\in-1/+1\}$,
number of nearest neighbour cells $n = 6$, $r = 1$ and we consider five different bcs.

\begin{figure}[ph]
\centerline{\includegraphics[width= 6.0cm]{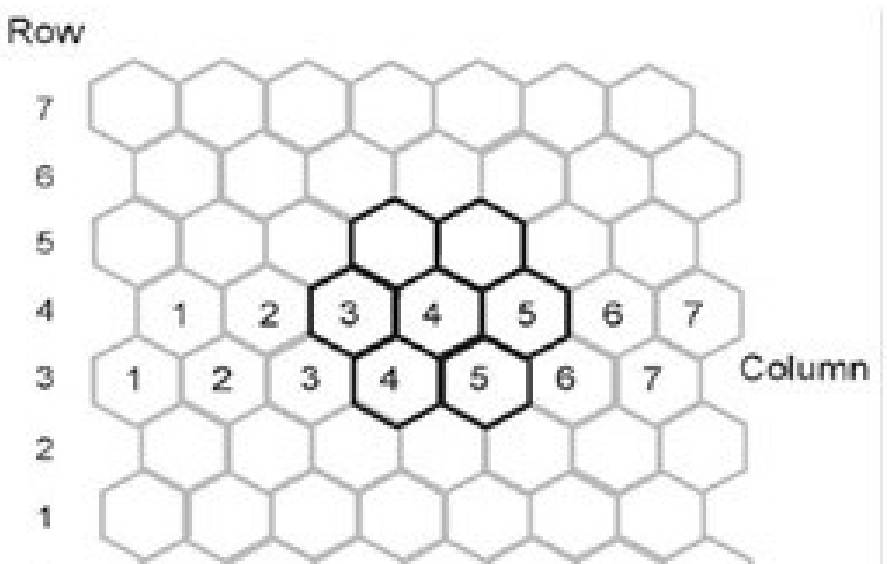}}
\vspace*{8pt}
\caption{Hexagonal-lattice with row and column vectors.\label{f1}}
\end{figure}
\begin{figure}[ph]
\centerline{\includegraphics[width= 6.0cm]{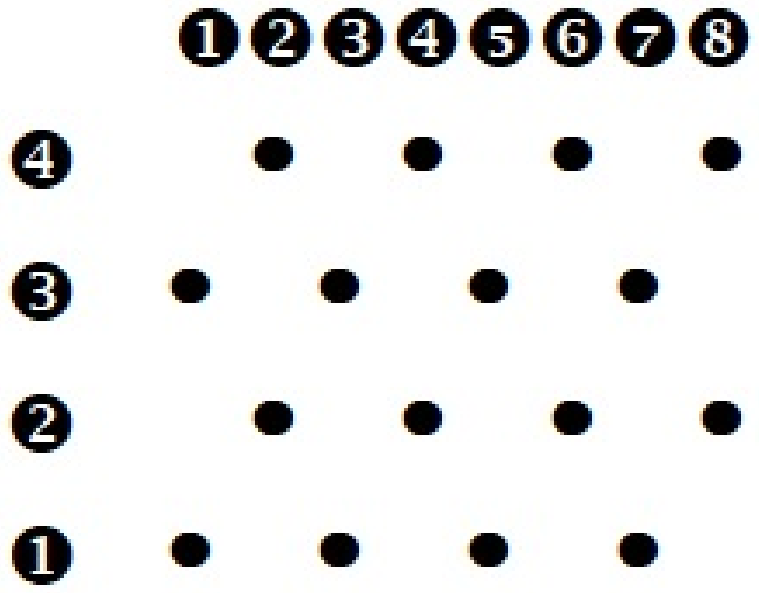}}
\vspace*{8pt}
\caption{$4\times4$ hexagonal-lattice.\label{f2}}
\end{figure}

\par Neighbourhoods of extreme cells are taken care of by boundary condition (bc).
If the extreme cells are adjacent to each other then it is called periodic bc (pbc).
In adiabatic bc (abc), the extreme cells replicate their state and in reflexive bc 
(rbc), mirror position states replace the extreme cells. In fixed bc, the extreme 
cells are connected to $+1$ or $-1$ state. If it is connected to $+1$ state, it is 
called fixed $+1$ bc ($f+1bc$) and if it is connected to $-1$ state, then it is 
called fixed -1 bc ($f-1bc$).

\par If the same rule is applied to all the elements of the matrix ($s$), then it 
is called \emph{uniform CA} and if different rules are applied to individual elements 
of the matrix or block of elements then it is called \emph{nonuniform CA}. At different 
time intervals, if different rules are applied to the matrix then it is called varying CA 
e.g., \emph{probabilistic CA}. With the application of these rules, elements (states) of 
the matrix change at successive intervals as shown in the following equation.

\begin{eqnarray}
s_{L \times L}^{t+1} = f_{L \times L}^{t} \times s_{L \times L}^{t} \label{eq:s_}
\end{eqnarray}
where $f$ is a time varying rule or transition matrix.

\par Consider an isotropic $2d$ Ising model in the form of hexagonal lattice
($s$) with $L$ rows and $L$ columns. Here we consider each hexagonal cell
as a lattice point as shown in figure 2. The lattice has then $L^2=N$ sites.
The odd rows have different column positions i.e., $1, 3, 5,\dots $ etc. and
even rows have different column positions i.e., $2, 4, 6,\dots$ etc.
At each of the sites $s_{i,j},\,i=1,\,\dots,\, L; j=1,\,\dots,\,2L$,
the spins are arranged in such a way that $j$ increases as $1, 3, \,\dots,\, 2L-1$
from left to right in case of odd rows, $j$ increases as $2, 4, \,\dots,\, 2L$
from left to right in case of even rows and $i$ increases from bottom to top,
and has one of the $\pm 1$ spin, which are the two states in CA. So, there are 
$2^{L^2}$ spin configurations. We consider only the nearest neighbor interactions, 
so the number of neighbor cells are $6$. We include the five different bcs as follows.

\emph{For odd rows :}
\begin{enumerate}
 \item pbc : $s_{i,2L+1} = s_{i,1}$, $s_{L+1,j+1} = s_{1,j}$,\\
             $s_{i,-1} = s_{i,2L-1}$ and $s_{-1,j} = s_{L,j+1}$.
 \item abc : $s_{i,2L+1} = s_{i,2L-1}$, $s_{L+1,j+1} = s_{L,j}$,\\
             $s_{i,-1} = s_{i,1}$ and $s_{-1,j} = s_{1,j-1}$.
 \item rbc : $s_{i,2L+1} = s_{i,2L-3}$, $s_{L+1,j+1} = s_{L-1,j+1}$,\\
             $s_{i,-1} = s_{i,3}$ and $s_{-1,j} = s_{2,j}$.
 \item f+1bc : $s_{i,2L+1} = +1$, $s_{L+1,j+1} = +1$,\\
              $s_{i,-1} = +1$ and $s_{-1,j} = +1$.
 \item f-1bc : $s_{i,2L+1} = -1$, $s_{L+1,j+1} = -1$,\\
               $s_{i,-1} = -1$ and $s_{-1,j} = -1$.
\end{enumerate}
\emph{For even rows :}
\begin{enumerate}
 \item pbc : $s_{i,2L+2} = s_{i,2}$, $s_{L+1,j} = s_{1,j-1}$,\\
             $s_{i,0} = s_{i,2L}$ and $s_{0,j} = s_{L,j+1}$.
 \item abc : $s_{i,2L+2} = s_{i,2L}$, $s_{L+1,j} = s_{L,j+1}$,\\
             $s_{i,0} = s_{i,2}$ and $s_{0,j} = s_{1,j-1}$.
 \item rbc : $s_{i,2L+2} = s_{i,2L-2}$, $s_{L+1,j} = s_{L-1,j}$,\\
             $s_{i,0} = s_{i,4}$ and $s_{0,j} = s_{2,j}$.
 \item f+1bc : $s_{i,2L+2} = +1$, $s_{L+1,j} = +1$,\\
              $s_{i,0} = +1$ and $s_{0,j} = +1$.
 \item f-1bc : $s_{i,2L+2} = -1$, $s_{L+1,j} = -1$,\\
               $s_{i,0} = -1$ and $s_{0,j} = -1$.
\end{enumerate}

\par The average magnetization for the configuration is defined as,

\begin{equation}
 \left\langle M \right\rangle = \sum\limits_{i=1}^{L} {\sum\limits_{\genfrac{}{}{0pt}{}{j=1}{i\%2=j\%2}}^{2L}} s_{i,j} \label{eq:M}
\end{equation}
and the average magnetization per spin is given by
\begin{equation}
  \left\langle m \right\rangle = \frac{\left\langle M \right\rangle}{N}
  \label{eq:m}
\end{equation}

\par Energy for the configuration $s$ is defined as
\begin{equation}
E(s) = - \frac{J}{6}\sum\limits_{i=1}^{L} {\sum\limits_{\genfrac{}{}{0pt}{}{j=1}{i\%2=j\%2}}^{2L}} s_{i,j}\times (s_{i,j-2}+s_{i,j+2}+s_{i-1,j-1}+s_{i-1,j+1}+s_{i+1,j-1}+s_{i+1,j+1})  \label{eq:E}
\end{equation}

Here, the factor of $1/6$ has been put to remove the sextuple
counting of energy, otherwise the interacting energy will be computed six times.
$J_{ij} = J$ (isotropic) for $6$ neighbour cells, or else, $J_{ij}=0$.

The configuration energy per spin is

\begin{equation}
\left\langle e \right\rangle  = \frac{E(s)}{N} \label{eq:e}  %
\end{equation}

\par For updating the lattice in next iteration, we use the probabilistic
approach by constructing a probabilistic CA. We use the following procedure.

\par We calculate the change in energy i.e., the energy difference at successive time intervals
is $\Delta E(s^t) = E(s^t) - E(s^{t-1})$. Here, $\Delta E \ge 0$ or $\Delta E < 0$ but
we have considered the case $\Delta E \ge 0$ i.e., $E(s^t)\ge E(s^{t-1})$. Next we calculate
the probability of each site in the spin configuration $s$ at time $t$ (number of iterations)
by using the Boltzmann factor

\begin{equation}
 p_t = \frac{p(E(s^t))}{p(E(s^{t-1}))} = e^{-\frac{\Delta E(s^t)}{k_BT}} \label{eq:p}
\end{equation}

With the above probability for each site, we construct a probability weighted matrix
(or transition matrix). This matrix leads to our probabilistic CA matrix ($PCA^t$)
by comparing it with a random matrix and multiplying by a factor of $0.1$
to normalise the $PCA^t$.

Successive spin configurations are obtained from

\begin{equation}
 [s_{i,j}^{t+1}]_{L\times L} = [PCA_{i,j}^t]_{L\times L} [s_{i,j}^{t}]_{L\times L} \label{eq:s_L}.
\end{equation}

Here, we consider $s_{i,j}^{t+1}$ to be a new configuration. If in this new
configuration the calculated  energy is less than  the initial configuration,
then it is allowed to proceed forward; otherwise the new configuration is
flipped backward. After a finite number of iterations we calculate the
average energy of the system per cell ($e$), magnetisation per cell ($m$),
and obtain the susceptibility per cell ($\chi$), specific heat per cell
($C_v$) and the entropy per cell ($S$).

\begin{equation}
 \chi = \frac{N}{k_B T}(\left\langle m^2 \right\rangle -
 \left\langle m \right\rangle ^2)  \label{eq:chi}
\end{equation}

\begin{equation}
 C_v = \frac{N}{k_B T^2}(\left\langle E^2 \right\rangle -
 \left\langle E \right\rangle^2)  \label{eq:C}
\end{equation}

\begin{equation}
 S = -k_B (r_1 P_1 \log_2 P_1 - r_2 P_2 \log_2 P_2)  \label{eq:S}
\end{equation}

where $r_1$ is the total number of spin up states, $r_2$ is the total number
of spin down states, $P_1$ is the probability of spin up states and $P_2$
is the probability of spin down states in the lattice $s$. The Metropolis algorithm
specifies that transitions must be made for one site at a time. But here we have
considered transitions of many sites simultaneously for which we go to a regime
of oscillating behavior. For instance, nearly every site that is flipped in the
direction of higher energy becomes unflipped on the next iteration.
CA is desirable to perform transitions on every lattice site simultaneously. For
instance such simultaneous transitions take advantage of matrix operations for
increased computational efficiency. This algorithm checks the time complexity better
than the Metropolis algorithm \cite{26}.

\section{Simulation, results and discussions}
\label{sec:simulation}
In this work, we have considered hexagonal-lattice of
size $40\times40$ for the determination of critical temperature
at which a phase transition occurs. Here we have considered lattice
size of $40\times40$ for our convenience in computation.
In this simulation  we take $J = 1$, $k_B = 1$ and the
external magnetic field to be zero i.e., $h=0$.
We have considered the temperature range from $0.1$ to $4.0$ with
increment of $0.05$ unit in each step. At each temperature, we have
considered $10^5$ runs. Each of the simulation starts with random
spin configurations. We have considered five different bcs i.e.,
periodic, adiabatic, reflexive, fixed $+1$ and fixed $-1$.

The magnetisation per cell ($m$) versus temperature results are shown for all
five bcs  in figure 3 . One can observe that in the case of fixed $\pm1$ bcs,
the curves obtained provide better result than either of the other boundary
conditions, i.e., pbc, abc or rbc at low temperature regions ($T<T^H_c$).

\begin{figure}[!ht]
\centerline{\includegraphics[width= 20cm]{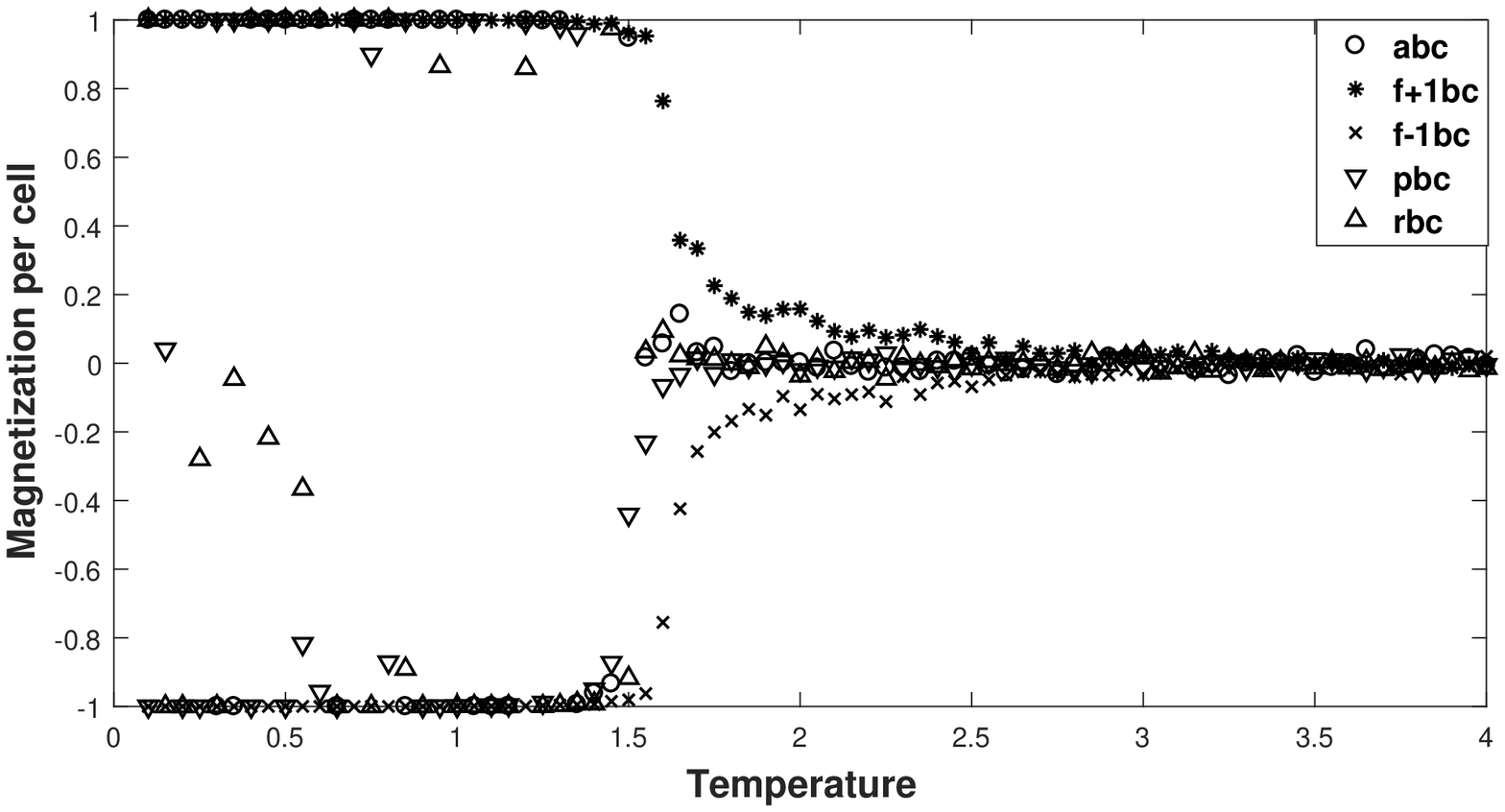}}
\caption{$m$ versus $T$ for all five bcs with random initial spin configuration in $25\times25$ hexagonal-lattice.\label{f3}}
\end{figure}

\subsection{Phase transition with pbc, abc, rbc, f1bc and f-1bc}
\par To study the phase transition we have calculated energy per cell ($e$),
magnetisation per cell ($m$), susceptibility per cell ($\chi$),
specific heat per cell ($C_v$) and entropy of the configuration ($S$)
by considering average of ten simulation. Here each simulation
starts with random spin configuration in hexagonal-lattice of
size $40\times40$ and we take temperature range from $1.0$ to $4.0$
with increment of $0.05$ unit since we have to find the phase
transition corresponding to all five different bcs and hence the $T^H_c$.

\par In figure 4, we have plotted $e$ vs $T$; $m$ vs $T$; $m$ vs $e$;
$\chi$ vs $T$; $C_v$ vs $T$ and $S$ vs $T$ with initial random spin configuration
with pbc, abc, rbc, f+1bc and f-1bc respectively. One finds that
around temperature $T^H_c=1.5$, the magnetisation approaches
the value zero, the energy gradually increases, as well as the
entropy gradually increases, as shown in figures 4(a), 4(b) and 4(f).
The susceptibility and specific heat also change, initially they
increase with a sudden rise at $T = 1.5$ and then start decreasing as shown in
figures  4(d) and 4(e) respectively. So, a phase transition is
clearly visible around $T^H_c = 1.5$ as the second order derivative
of energy and magnetisation is extremely large with all five bcs.

\par In the $m$ vs $e$ graphs shown in figure 4(c), the regions with
higher density of points indicate three states. We find three states,
two of them are low temperature ground states around ($m = \pm1$, $e = -6$)
with random initial conditions in the cases of abc, pbc and rbc, and the third one
is the high temperature phase which is centered at ($m = 0$, $e = 0$).
But in the case of f+1bc there is one low temperature ground state around
($m = 1$, $e = -6$) and one high temperature phase is centered at
($m = 0$, $e = 0$). Also in the case of f-1bc there is one low temperature
ground state around ($m = -1$, $e = -6$) and one high temperature phase is
centered at ($m = 0$, $e = 0$).

\begin{figure}[!ht]
  \begin{tabular}{cc}
  \includegraphics[width=7cm]{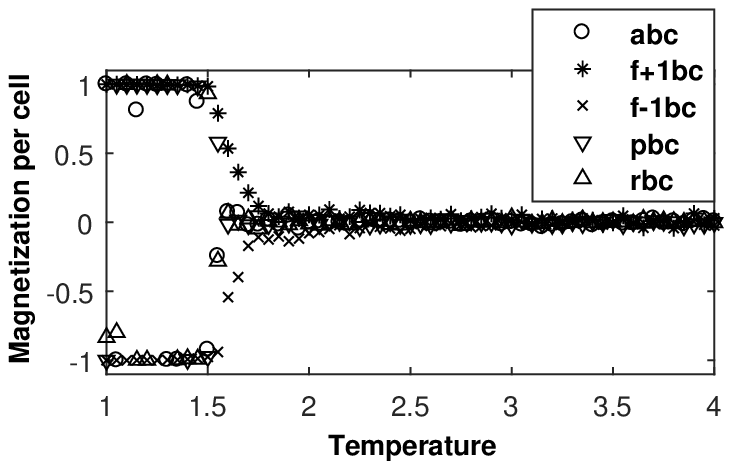} &   \includegraphics[width=7cm]{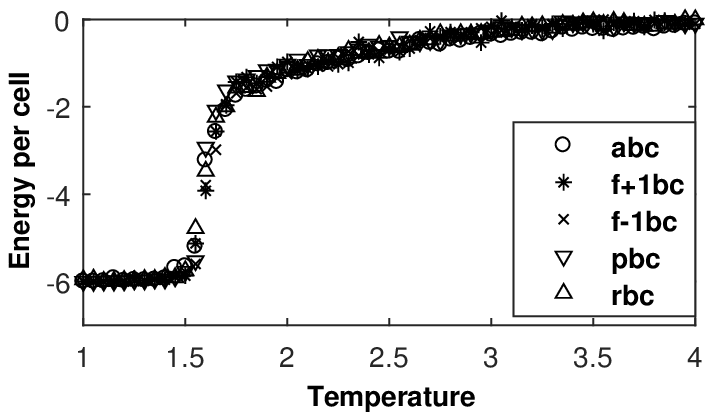}
  \\(a) $m$ versus $T$  & (b) $e$ versus $T$ \\
  \includegraphics[width=7cm]{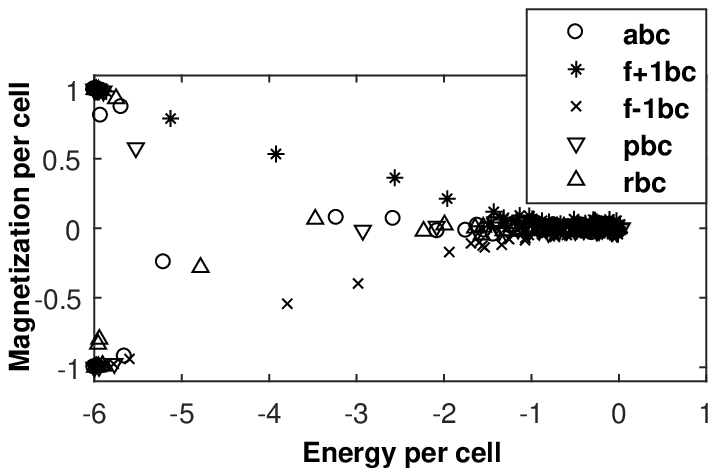} &   \includegraphics[width=7cm]{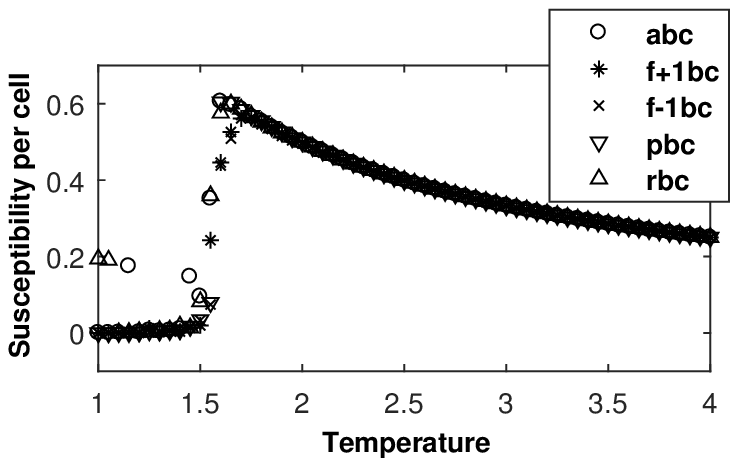}
  \\(c) $m$ versus $e$  & (d) $\chi$ versus $T$ \\
  \includegraphics[width=7cm]{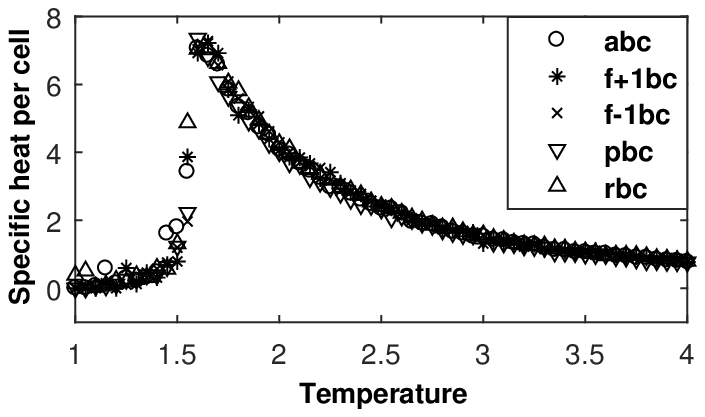} &   \includegraphics[width=7cm]{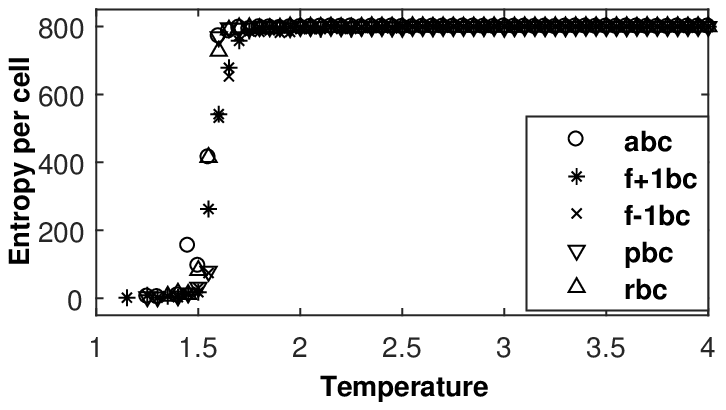}
  \\(e) $C_v$ versus $T$ & (f) $S$ versus $T$
  \end{tabular}
  \caption{Random initial spin configuration in $40\times40$ hexagonal-lattice with five different bcs.\label{f4}}
\end{figure}

\section{Comparison of $2d$ Ising model simulation among boundary conditions}
\label{sec:comparision}
\par In figure 3, one can observe that the magnetization approach the zero line after $T > T^H_c$
in different manner depending on the bcs. The analytic solution gives a zero magnetization above
$T^H_c$ and our simulation for different bcs also show that it converge toward the zero
value at increased temperatures. With one simulation for all bcs, it is not possible
to predict which bc is closer to $T^H_c$. So, we analyze the points for magnetization in the
range $-0.01 \leq m \leq +0.01$, $-0.1 \leq m \leq +0.1$ and $-0.2 \leq m \leq +0.2$
which are close to the zero line of magnetization after $T > T^H_c$. We call such points 
as converging points.

\par In this simulation, we have taken various lattice sizes ranging from $5\times5$ to 
$200\times200$ in this manner. We consider an increment of lattice size $5$ in each step 
while going from lattice size from $5\times5$ to $60\times60$ and subsequently for 
$60\times60$ to $100\times100$ an increment of $10$ for each step and from $100\times100$ 
to $200\times200$ with increment of $50$ for each step. The temperature ranges from 
$1.0$ to $4.0$ with small increments of $0.05$ units. We have iterated $1000$ times 
for each  simulation for lattice sizes $\leq 10\times10$, $10000$ times for lattice 
sizes in between $10\times10$ and $40\times40$ and $100000$ times for lattice sizes 
above $40\times40$. Figure 5 shows the converging points in the above mentioned range 
of magnetization. We have counted the number of converging points as defined above by 
taking average result of ten simulations for each bc with three different initial conditions.
We carry out all the simulation with three initial conditions and with all five bcs.
Here, three initial conditions are (i) all up (or most of spins up), (ii) all down 
(or most of the spins down) and (iii) random (or randomly oriented spins up/down).

\begin{figure}[!ht]
\centerline{\includegraphics[width= 20cm]{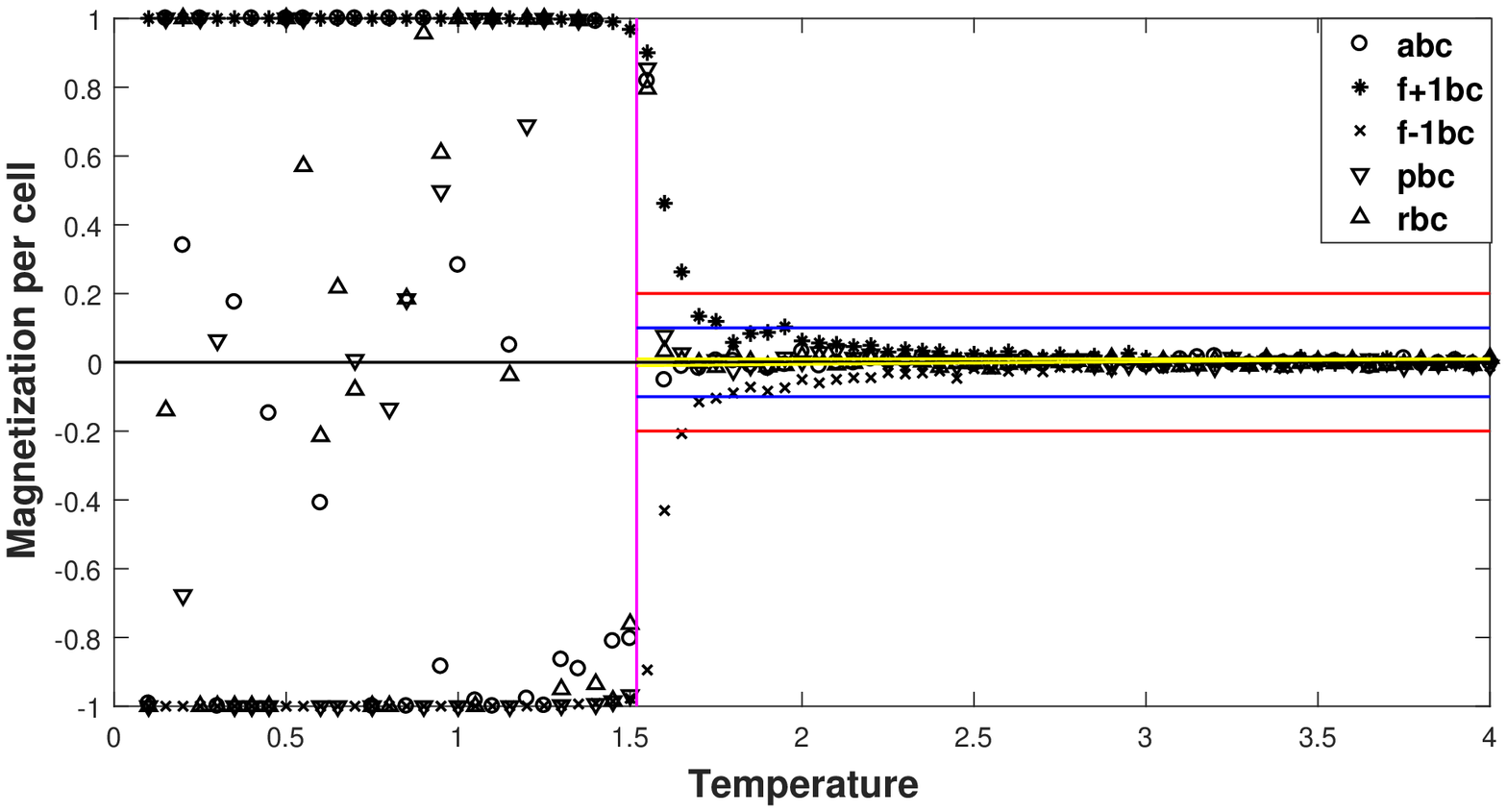}}
\caption{Converging points of bcs for $50\times50$ lattice size after $T > T^H_c$.
Red lines are for magnetization $|m|\leq0.2$, blue lines are for magnetization
$|m|\leq0.1$ and yellow lines are for magnetization $|m|\leq0.01$ with
the initial condition of random spin configuration. Magenta line represents
a parallel line to magnetization per cell at $T = T^H_c$.\label{f5}}
\end{figure}

\par For lattice $sizes \le 50\times50$ with pbc, abc and rbc, one finds more convergent points
in both cases $-0.1 \le m \le +0.1$ and $-0.2 \le m \le +0.2$ than that for f+1bc
and f-1bc for all up and all down spin initial conditions.

\par With lattice sizes in between $50\times50$ and $200\times200$ with f-1bc, one finds more
convergent points in both the cases corresponding to $-0.1 \le m \le +0.1$ and $-0.2 \le m \le +0.2$
with initial condition of all up spins. For f+1bc, one finds more convergent points in both
the cases with initial condition of all spins down.

\par In the case of lattice $sizes \leq 30\times30$ with pbc and lattice $size > 30\times30$
with abc, one finds more convergent points both cases $-0.1 \le m \le +0.1$
and $-0.2 \le m \le 0.2$ among all bcs with random spin initial conditions.
For all initial cases with $-0.01 \le m \le +0.01$, the result is better for pbc,
abc and rbc and converging points increase as lattice sizes are increased.

\par It is observed that lattice $size = 200\times200$ with all five bcs, one can observes 
same convergent points in both cases $-0.1 \le m \le +0.1$ and $-0.2 \le m \le +0.2$ but pbc,
abc and rbc show more convergent points in $-0.01 \le m \le +0.01$ than f+1bc and f-1bc.

\section{Conclusion}
\label{sec:conclusion}
\par In the present work, we have studied $2d$ Ising model in hexagonal-lattice
with five different boundary conditions using non-deterministic Cellular Automata
algorithm in which transitions of many sites takes place simultaneously in contrast 
to transitions being made for one site at a time. We have observed a phase transition 
that occurs at the critical temperature $T^H_c\approx1.5$ for each of the five bcs. 
The result of $T^H_c\approx1.5$ agrees with the exact solution presented in equation $1.3$. 
The findings for fixed $+1/-1$ bcs also show a smoother curve at low temperatures than 
pbc, abc and rbc with random initial condition. This implies that with the different 
initial conditions on different lattice sizes $\le 50\times50$, one can take care of 
boundary spins by not only pbc but also by abc and rbc. Further in our analysis, abc 
shows more converging points than pbc and rbc for lattice size $> 30\times30$ with 
random initial condition. For lattice size greater than $50\times50$, f+1bc and
f-1bc are better suited to use than other bcs in case of initial spin configuration 
with all down or all up spins respectively. It is also observed that for lattice 
$size = 200\times200$, in all cases of initial spin configuration, in all bcs give 
almost same converging points. From the simulation point of view, our algorithm
flipped most of the spins at a single iteration. It is computationally more 
efficient than Metropolis algorithm \cite{26} which flips single spins at a single 
iteration. The analysis done in this paper will help us in finding the values of 
critical exponents more accurately for hexagonal-lattice, which we plan to study next. 
The present work is expected to facilitate the use of non-deterministic CA in the study 
of phase transition in case of anti-ferromagnetic materials, binary alloys and spin 
glasses etc with different boundary conditions.

\subsection*{Acknowledgement}
We thank K. Maharana and S. Pattanayak for useful discussions. JM would
like to thank Institute of Physics, Bhubaneswar for hospitality where
part of this work has been done.


\end{document}